\documentclass{amsart}
\pdfoutput=1

\usepackage{amsmath}
\usepackage{amssymb}
\usepackage{tikz}
\usepackage{subfig}

\newcommand{\FF}{\mathbb{F}}
\newcommand{\kk}{\FF_q}
\newcommand{\ZZ}{\mathbb{Z}}

\newcommand{\bits}[1]{{\tt #1}}

\begin{document}

\title{Bitslicing and the Method of Four Russians Over Larger Finite Fields}
\author{Tomas J. Boothby}
\author{Robert W. Bradshaw}

\date{\today}

\begin{abstract}
We present a method of computing with matrices over very small finite fields of size larger than 2.
Specifically, we show how the Method of Four Russians can be efficiently adapted to these larger fields, 
and introduce a row-wise matrix compression scheme that both reduces memory requirements and allows one to vectorize element operations. 
We also present timings which confirm the efficiency of these methods and exceed the speed of the fastest implementations the authors are aware of. 
\end{abstract}


\maketitle

\section{Introduction} \label{sec:intro}

Using the special properties of the finite field $\FF_2$ and the binary-based nature of modern computing, a wealth of specialized algorithms and optimized implementations have been developed 
for doing linear algebra over $\FF_2$. 
On the other hand, much work has gone into creating fast linear algebra over word sized primes as a basic building block of multi-modualar and $p$-adic methods. 
In this paper first we present a method of computing with matrices over finite fields that are significantly smaller than a single machine word, but larger than $\FF_2$. 
Such matrices arise for example in number theory \cite{mod5modularity} and graph theory \cite{MaySauWan07, WengQuiWangXiang07}. 

We show how the Method of Four Russians can be efficiently adapted to finite fields larger than $\FF_2$, 
and introduce a row-wise matrix compression scheme that both reduces memory requirements and allows one to vectorize element operations. 
As row addition is the essential operation in the method of the four Russians, these two techniques 
go very well together. 
We demonstrate the practicality of these methods with timings. 

In section \ref{sec:m4rm} we present the Method of Four Russians for multiplication of matrices, and show how it can be used for the fields in question. 
In section \ref{sec:bitslice} we show how the idea of bitslicing yields a convenient packed representation,
and compare it to the representation used for very small prime fields in \cite{DuFoSa08}. 
In section \ref{sec:large} we give the specific representations used with justification, and timings are given in section \ref{sec:results}.

\subsection*{Acknowledgment}
The authors would like to thank Martin Albrecht, Joel Barnes, Gregory Bard, Bill Hart, and especially Cl\'ement Pernet and William Stein for helpful discussions and comments.  We are grateful to Bill Hart and the University of Warwick for access to the hardware used for benchmarking.

\section{Method of Four Russians} \label{sec:m4rm}

The Method of Four Russians (M4RM) was first introduced by Arlazarov, Dinic, Kronrod, and Faradzev
in the context of Graph theory \cite{AUH, M4RM} 
and has traditionally been used for boolean matrices. 
It has a runtime complexity of $O(n^3/\log n$), and has been extended to system solving and matrix inversion in additino to matrix multiplication \cite{Bard06}. 
Though this has worse asymptotic complexity than Strassen-Winograd and other lower-exponent matrix multiplication algorithms, the actual cutoff is often high enough to make it competitive for medium-sized problems \cite{Bard06}. It can also effectively be used as a base-case for asymptotically faster algorithms \cite{M4RI-paper}.

Consider the product of two matrices $C=AB$ where $A$ is an $m \times l$ matrix 
and $B$ is an $l \times n$ matrix. Let $C_i$ denote the $i$-th row of $C$. 
Then the rows of $C$ can be viewed as linear combinations of the rows in $B$, 
with coefficients selected according to the $i$-th row of A. That is, 
\[
C_i = A_iB = \sum_{j=0}^{l-1} a_{ij} B_j.
\]
Let $k$ be a small integer and, for simplicity, assume for the moment that $k$ divides $l$. We can write this sum as 
\[
C_i =  \sum_{j=0}^{l-1} a_{ij} B_j = \sum_{s=0}^{l/k-1} \;\; \sum_{t=0}^{k-1} a_{i(sk+t)} B_{sk+t}.
\]
When the field of definition $K$ is small, precomputations can be used to speed up the sums 
$\sum_{t=0}^{t-1} a_t B_{sk+t}$ for all $(a_0, ..., a_t) \in K^k$, which can be shared for all rows $C_i$. 
When $K = \FF_2$ this is done by creating lookup tables of binary combinations of the rows. 
Obviously, precomputing all possible linear combinations of the rows has diminishing returns 
when the cardinality of the field is any larger than two, so we adapt the method as follows.

Let $\alpha_0, ..., \alpha_r$ be an additive basis for $K$, with maps $\phi_0, ..., \phi_r : K \rightarrow \{0,1\}$ such that $a = \sum_{d=0}^r \alpha_d \phi_d(a)$ for all $a \in K$. 
Then we can write
\[ \label{eqn:m4rm_with_split}
\sum_{t=0}^{t-1} a_t B_{sk+t} 
= \sum_{t=0}^{t-1} \left(\sum_{d=0}^r \alpha_d \phi_d(a_t) \right) B_{sk+t} 
= \sum_{d=0}^r \alpha_d \sum_{t=0}^{t-1} \phi_d(a_t) B_{sk+t} 
\]
where the inner sum is now over the \emph{binary} combinations of rows of $B$, 
which is much more amenable to precomputation. 
One also has some freedom in choosing the additive basis such that the 
scalar products are easy to compute, for example letting 1 and -1 be in the basis, 
or choosing a power basis so Horner's rule can be applied.

\section{Bitslicing and Matrix Compression} \label{sec:bitslice}

The the most basic unit of arithmetic on a modern processor is a multi-bit word, 
and operating on individual bits usually cannot be done more cheaply (in fact, sometimes it's more expensive). 
However, if one is doing the same operations on many values, a standard trick is 
to pack multiple values into a single word and then do word-sized arithmetic on the packed values. 

For example, one might want add vectors over $\ZZ/4\ZZ$ by packing every element into two bits,
\[
 \langle 3,0,1,2,0,1,3,2 \rangle = \bits{11~00~01~10~00~01~11~10}.
\]
One would like to add two vectors of this form with standard integer addition, but of course, this does not work---instead, one must pack every element into three bits
\[
 \langle 3,0,1,2,0,1,3,2 \rangle = \bits{011~000~001~010~000~001~011~010}
\]
so we can handle the overflow. Addition is now normal integer addition and removing the cary bit (reduction mod 4). 
\begin{eqnarray*}
 c &\leftarrow& a+b\\
 d &\leftarrow& c \wedge \bits{011 ~ 011 ~\cdots~ 011}
\end{eqnarray*}
A priori, this looks pretty good: 21 element additions in 2 operations on a 64-bit processor.  But every third bit is wasted,
and if we wanted to perform scalar multiplication, even more padding would be needed. 
If only we could define our own arithmetic which ignores the carry, we'd be in business.  This is done by bitslicing.

In cryptography, for example, bitslicing has been used to speed up the computation of $S$-boxes in the DES cypher \cite{Biham97, Pornin99, Kwan00}.
Rather than storing the bits of a single element as adjacent bits in a single word, we store them as parallel bits in multiple words. So, we represent a vector of elements over $\ZZ/4\ZZ$ as a pair of words,
\[
 \langle 0,0,1,2,0,1,2,2 \rangle = \begin{array}{c} \bits{00010011}\\ \bits{00100100} \end{array}.
\]
Addition mod 4 of a pair of two-bit numbers $a_1a_0$ and $b_1b_0$ can be done with four bit operations
\begin{eqnarray*}
 r_0 &\leftarrow& a_0 \oplus b_0 \\
 r_1 &\leftarrow& (a_1 \oplus b_1) \oplus (a_0 \wedge b_0).
\end{eqnarray*}
If we view the inputs as machine words rather than individual bits, and perform bitwise operations on words, the addition formula holds at each bit of the inputs. 
One a 64-bit machine we can now add 64 elements with 4 operations, or 16 elements per instruction; whereas above, we only add 10.5 per instruction. 
Any operation that can be expressed in terms of boolean formulas can be vectorized in this way. 

The classical packing method is used in \cite{DuFoSa08} where multiple matrix entries into a single double-precicion floating point values and using optimized numerical linear algebra routines in the spirit of FFLAS/FFPACK \cite{FFLAS}. 
Simultaneous Modular Reduction \cite{REDQ} is used to perform the modular reductions. 
This has the advantage that one can leverage the highly-optimized and tuned floating point packages such as ATLAS, as well as getting multi-core or hardware acceleration for free if the underlying BLAS is compiled to take advantage of it. 
Unfortunately the matrix dimension and amount of padding needed for a dot product puts a rather severe upper bound on the number of entries that can be packed in a double. 
Specifically, at least $\log_2 n(p-1)^2$ bits need to be used per field element to compute a dot product of length $n$. This means, assuming a 53-bit mantissa, only $3$ entries could be stored per double when multiplying a $1000\times 1000$ matrix over $\FF_5$ or $\FF_7$. 
This packing scheme also has the disadvantage that left operand, right operand, or product are stored using different compression schemes.

\section{Arithmetic over Specific Fields} \label{sec:large}

To actually implement the algorithm for a specific field, 
one needs to find {\em short} boolean formulas which express the arithmetic in the field. 
More accurately, we are interested \textit{sequential 
program on $n$ inputs}, that is, a sequence of tuples,
\[
 (*_0,\{i_0,j_0\}), (*_1,\{i_1,j_1\}), \ldots, (*_\ell,\{i_\ell,j_\ell\})
\]
where each $*_k$ is any of $(\wedge, \vee, \oplus)$, and $-n \leq i_k < j_k < k$.  
Then, we can evaluate such a program by the recurrence
\[
 v_k \leftarrow v_{i_k} *_k v_{j_k},
\]
where $v_{-n}, \ldots, v_{-1}$ are the inputs to the program.  
Trying to find small programs by hand is a fun exercise, but it is difficult to prove such programs minimal. 

The number of the sequential programs on $n$ inputs with length $\ell$ is given by
\[
 N(n,\ell) = 3^\ell \prod_{k=0}^{\ell-1} \binom{k+n}{2}.
\]
For two-bit fields, a binary arithmetic operation has 4 inputs.  From the na\"ive 
count above, there are more than 128 million such programs of length 5, and over 13 
billion of length 6. This is still within the reach of an exhaustive computer search which we have performed.
For larger fields, the number of inputs is six or more, and the expected minimal program length is larger as well, so brute force searching methods seem to be prohibitively expensive. 

We note that this is equivalent to boolean logic and circuit minimization, and 
a considerable amount of reserach has been gone into optimizing such things. 
Unfortunately, the standard methods such as Karnaugh maps\cite{Karnaugh} and 
the Espresso algorithm \cite{Espresso} performed very poorly on these particular circuits and 
often produced far from optimal results.  For example, in our representation 
of $\FF_3$, Logic Friday \cite{LogicFriday} (which implements the Espresso algorithm) produces circuits 
with 12 or more gates---twice as large what is actually required.

\subsection{Arithmetic over $\FF_3$} \label{sec:gf_three}

The application of the Method of Four Russians to non-binary fields began, naturally, 
with an investigation into its feasibility over $\FF_3$.  
We use the additive basis $\{1, -1\}$. 
One may be tempted to use the binary representation
\[
 0 = \bits{00}, 1 = \bits{01}, -1 = \bits{10},
\]
but we find that the minimal addition requires 7 operations per word in this representation.
Instead, we use the representation
\[
 0 = \bits{00}, 1 = \bits{10}, -1 = \bits{11},
\]
so the first bit $x_0$ marks units, and the second bit $x_1$ indicates the 
sign of the element.  In this representation, vector addition requres only 6 bitwise 
operations per pair of words,
\begin{eqnarray*}
s &\leftarrow& x_0 \oplus y_1 \oplus x_1 \\
t &\leftarrow& x_1 \oplus y_0 \oplus y_1 \\
r_0 &\leftarrow& (x_0 \oplus y_1) \wedge (x_1 \oplus y_0)\\
r_1 &\leftarrow& s \vee t.
\end{eqnarray*}

Next, we note that negation requires one operation,
\begin{eqnarray*}
 r_0, r_1 &\leftarrow& a_0, a_0 \oplus a_1
\end{eqnarray*}
so we can automatically perform vector subtraction in 7 operations.  However, we can do 
one better:
\begin{eqnarray*}
t &\leftarrow&  x_0 \oplus y_0 \\
r_0 &\leftarrow&  t \vee   (x_1 \oplus y_1)\\
r_1 &\leftarrow&  (t \oplus y_1) \wedge (y_0 \oplus x_1).\\
\end{eqnarray*}

We compare this to the classical packing method, in which each element is packed into three bits of a word.  Here,
a row sum is computed by
\begin{eqnarray*}
 x &\leftarrow& (a + b) \wedge \bits{011 ~ 011 ~ \cdots ~ 011}\\
 y &\leftarrow& (a + b) \wedge \bits{100 ~ 100 ~ \cdots ~ 100}\\
 r &\leftarrow& x + \frac{1}{4}y
\end{eqnarray*}
where the division in the last step is performed via a bit shift. 
For a 64-bit word, we perform 21 element additions in 5 operations, compared to 64 additions in 6 operations.  In this case bitslicing more than doubles the speed of computation over classical integer packing.

\subsection{Arithmetic over $\FF_5$ and $\FF_7$} \label{sec:gf_five_seven}

For $\FF_3$, we were able to find representatives for the field elements giving nice arithmetic formulas.
Such representations for larger fields, if they exist at all, are quite elusive. 
However, the standard binary representation works fairly well if one relaxes the requirement that representations be unique. 
The additive basis we choose in this case is $\{1, 2, 4\}$ which corresponds nicely with our representation. Now to use the multiplication algorithm specified above, one only needs to specify how to add and double elements using only bit operations. 
For completeness, it is useful to be able to negate and reduce to a canonical representative. 

Denote the standard grade-school addition on two binary integers by {\tt add}. 
For an $n$ and $m$-bit input, and without loss of generality assuming $m \le n$, 
this can be done with $m-1$ full adders and $n-m+1$ half adders using a total 
of $5(m-1) + 2(n-m+1) = 3m+2n-3$ bit operations. 

For $\FF_7$ the ``carry'' bit from standard 3-bit addition is equal to the unit bit mod 7. This gives particularly nice formulas. 

\begin{itemize}
\item Addition (17 bit operations):
\begin{eqnarray*}
s_3s_2s_1s_0 &\leftarrow& {\tt add}(a_2a_1a_0, b_2b_1b_0) \\
r_2r_1r_0 & \leftarrow & {\tt add}(s_2s_1s_0, s_3) \\
\end{eqnarray*}
It is easy to see there will not be a carry in the last {\tt add} as not all of $s_0, ..., s_3$ can be 1. 

\item Double (0 bit operations):
\begin{eqnarray*}
r_0, r_1, r_2 &\leftarrow& a_2, a_0, a_1
\end{eqnarray*}

\item Negate (3 bit operations):
\begin{eqnarray*}
r_0, r_1, r_2 &\leftarrow& \overline{a_0}, \overline{a_1}, \overline{a_2}
\end{eqnarray*}

\item Reduce (5 bit operations):
\begin{eqnarray*}
r_0 &\leftarrow& a_0 \oplus (a_0 \vee a_1 \vee a_2) \\
r_1 &\leftarrow& a_1 \oplus (a_0 \vee a_1 \vee a_2) \\
r_2 &\leftarrow& a_2 \oplus (a_0 \vee a_1 \vee a_2)
\end{eqnarray*}

\end{itemize}
Clearly the formulas for $\FF_7$ generalize to a general $n$-bit Mersenne prime.

Things aren't as nice for $\FF_5$, but one can still find acceptable formulas. We introduce an auxiliary function ${\tt fold5}$ which takes a 4-bit input $s_3s_2s_1s_0$ and ``folds'' the highest bit into the other three, preserving the value mod 5. 

\begin{itemize}
\item {\tt fold5}:

The best comprehensible formula we were able to come up with has 13 operations.
\begin{eqnarray*}
n_0, n_1, n_2 & \leftarrow & \overline{s_2}, s_3, \overline{s_3} \\
e_3e_2e_1e_0 & \leftarrow & {\tt add}(n_2n_1n_0, s_1s_0) \\
r_0, r_1, r_2 & \leftarrow & e_0 \wedge e_3, e_1 \wedge e_3, e_2
\end{eqnarray*}

Using a computer-assisted search, we found a shorter (8 operation) but entirely cryptic formula:
\begin{eqnarray*}
    t & \leftarrow & s_2 \vee s_1 \\
    r_2 & \leftarrow & s_0 \oplus \,t \\
    r_1 & \leftarrow & (r_2 \wedge s_0) \oplus (s_3 \oplus s_1) \\
    r_0 & \leftarrow & (\;t\; \oplus s_2) \vee (r_1 \wedge s_3)
\end{eqnarray*}

\item Addition (20 bit operations):
\begin{eqnarray*}
s_3s_2s_1s_0 &\leftarrow& {\tt add}(a_2a_1a_0, b_2b_1b_0) \\
r_0, r_1, r_2 & \leftarrow & {\tt fold5}(s_3s_2s_1s_0)
\end{eqnarray*}

\item Double (5 bit operations):
\begin{eqnarray*}
r_0, r_1, r_2 &\leftarrow& {\tt fold5}(a_2a_1a_0\bits{0})
\end{eqnarray*}

\item Negate (6 bit operations):
\begin{eqnarray*}
r_0, r_1, r_2 &\leftarrow& {\tt fold5}(a_1a_0\bits{0}a_2)
\end{eqnarray*}

\item Reduce (6 bit operations):
\begin{eqnarray*}
t & \leftarrow & a_0 \oplus a_1 \\
u & \leftarrow & (a_0 \wedge a_1) \vee t \\
r_0, r_1, r_2 & \leftarrow & u, (u \vee a_2) \oplus a_0, u \oplus t
\end{eqnarray*}

\end{itemize}

These approaches do not seem to adapt themselves well to primes of a more general form. 

Though the rings $\ZZ/2^k\ZZ$ typically aren't very interesting rings to do linear algebra over, it is worth noting that these rings lend themselves to very short formulas of the form above by simply ignoring the last carry bit. 
For example, in $\ZZ/8\ZZ$ one gets an 11 operation add, a 0 operation double, and a 7 operation negate, with the benefit that the representation is unique. 

\subsection{Non-prime fields}

Similar methods can be applied to extension fields, and it is trivial to come up with particularly nice formulas for $\FF_{2^2}$, $\FF_{2^3}$, etc.
For a general $\FF_{p^n}$, elements can be added by $n$ repeated applications of the addition formula for $\FF_p$, and the additive basis can be chosen to be all products of the power basis of $\FF_{p^n}$ with the additive basis if $\FF_p$ which is $n$ times as large. 
This allows one to perform multiplication in $n^2 + o(n^2)$ times the number of bit operations needed to multiply over the base field. 
However, even more substantial gains can be achieved by considering bitslicing at the level of matrices rather than rows. Consider the matrices 
\[
A =  A_0 + \alpha A_1 ~~ \mbox{and} ~~ B = B_0 + \alpha B_1
\]
Where $A_0, A_1, B_0, B_1$ are over $\FF_2$ and $\alpha$ is a generator of $\FF_{2^2}$. Using Karatsuba and reducing modulo $x^2 + x + 1$, one can compute their product as 
\[
AB 
= (A_0 B_0 + A_1 B_1) + \alpha \left((A_0 + A_1)(B_0 + B_1) + A_0 B_0\right).
\]
This requires only three matrix multiplies over $\FF_2$, a significant advantage.  
In general one can view a matrix $A$ over $\FF_{p^n}$ as
 \[
 A = A_0 + \alpha  A_1 + \cdots + \alpha^{n-1}A_{n-1}
 \]
 where each $A_i$ is a matrix over $\FF_p$. 
One can then use fast polynomial multiplication techniques to reduce the number of matrix multiplications for a product of two matrices of this form, and reduction by the defining polynomial only involves addition and possibly multiplication by a scalar.

Unfortunately Toom-Cook multiplication requires more distinct elements than may be available in the base field, but (potentially repeated use of) Karatsuba works in any field, and trinomials $a$ and $b$ can be multiplied in any field with 6 coefficient multiplies using the Karatsuba-like formula
\begin{eqnarray*}
c_0 &=& a_0b_0 \\
c_1 &=& a_0b_1 + a_1b_0 = (a_0+a_1)(b_0+b_1) - a_0b_0 - a_1b_1 \\
c_2 &=& a_0b_2 + a_1b_1 + a_2b_0 = (a_0+a_2)(b_0+b_2) - a_0b_0 - a_2b_2 + a_1b_1 \\
c_3 &=& a_2b_1 + a_1b_2 = (a_2+a_1)(b_2+b_1) - a_2b_2 - a_1b_1 \\
c_4 &=& a_2b_2.
\end{eqnarray*}
which is significantly better than the 9 multiplies using elementary polynomial multiplication, so provides an advantage for cubic extension fields.
Over fields with 5 or more elements, Toom-3 multiplies trinomials with 5 coefficient multiplies.  
When the matrix dimensions are much larger than the degree of the extension and the base field has enough elements, the large number of additions in higher-degree Toom-Cook algorithms can still be offset by saving a single matrix multiply. 
There is likely to be an additional constant speedup as the elements manipulated in the innermost loops of the linear algebra routines are algebraically simpler, and the smaller footprint of the matrix entries results in better memory locality across the matrix. 
Further, this enables one to leverage optimized base field code for extension fields instead of writing extensive amounts of code from scratch, or overly generalizing the code used to compute linear algebra over small prime fields. 
This may also be useful for doing arithmetic with matrices over number fields.

\section{Implementation and Timings} \label{sec:results}

\begin{table}[h]
\centering
\begin{tabular}{l|rr|rr|rr|rr|rr|}
&
\multicolumn{2}{|c|}{100} & 
\multicolumn{2}{|c|}{500} & 
\multicolumn{2}{|c|}{1000} & 
\multicolumn{2}{|c|}{2500} \\
&
\multicolumn{1}{|c}{~B.}&\multicolumn{1}{c|}{~M.}&
\multicolumn{1}{|c}{~B.}&\multicolumn{1}{c|}{~M.}&
\multicolumn{1}{|c}{~B.}&\multicolumn{1}{c|}{~M.}&
\multicolumn{1}{|c}{~B.}&\multicolumn{1}{c|}{~M.}\\
\hline
$\FF_3$     & 
0.032	&	0.047 &
1.68	&	2.91& 
12.2	&	21.4&
199	&	266 \\
\hline
$\FF_5$     & 
0.110   & 0.143 & 
6.47   & 8.62    & 
49.4  & 62.7    & 
742     & 848     \\
\hline
$\FF_7$     & 
0.105   & 0.181 & 
6.04  & 10.9   & 
45.7   & 79.3    & 
672     & 1070   \\
\hline
$\FF_{2^2}$ & 
0.091   & 0.037 & 
1.89  & 2.13   & 
9.5  & 15.1  & 
132     & 203     \\
\hline
$\FF_{2^3}$ & 
0.187   & 0.101 & 
3.85  & 5.50    & 
20.6  & 40.1    & 
261    & 499     \\
\hline
$\FF_{3^2}$ & 
0.097   & 0.842 & 
5.22  & 62.0     & 
37.6  & 444.0   & 
601     & 6700    \\
\hline
\end{tabular}
\caption{Time to multiply $n$-dimensional square matrices over $\kk$ in milliseconds. Author Bitslicing implementations vs. Magma V2.15-3 on a 2.6GHz Opteron machine. }
\label{tab:bitslice-magma}
\end{table}


We have implemented matrix multiplication methods over $\FF_3, \FF_5$ and $\FF_7$, as well as quadratic and cubic extensions of these fields, using the ideas presented above.  
We also implemented $\FF_{2^2}$ and $\FF_{2^3}$ using the M4RI library \cite{m4ri} for arithmetic over $\FF_2$. 
In each case, our implementations are nearly always
faster than Magma \cite{magma} whose finite field linear algebra are the fastest known to the authors (see table \ref{tab:bitslice-magma}). 

\begin{figure}[h]
 \subfloat[$\FF_{3}$]
 {
 \includegraphics[scale=.4]{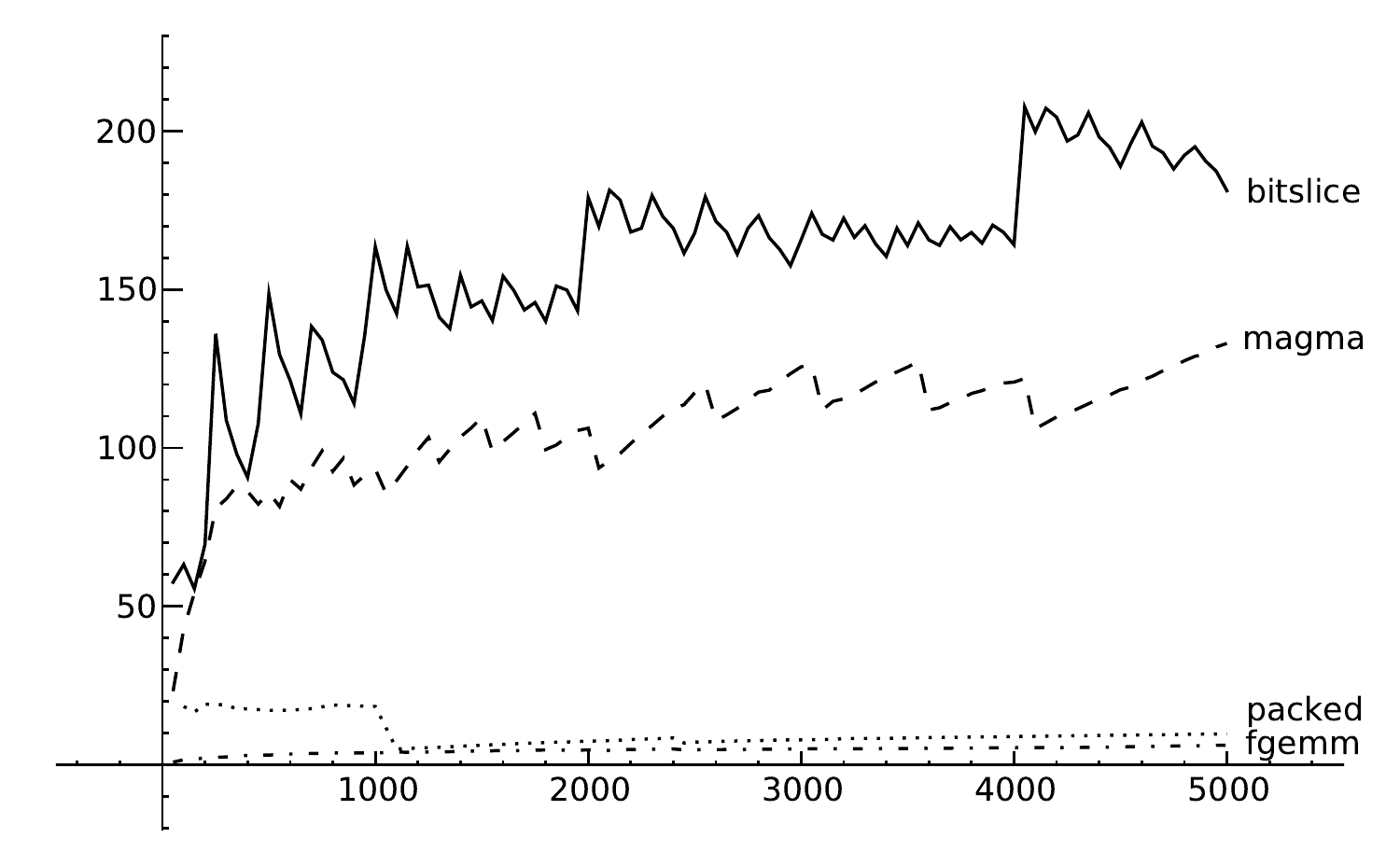}
 }
 \subfloat[$\FF_{2^3}$]
 {
 \includegraphics[scale=.4]{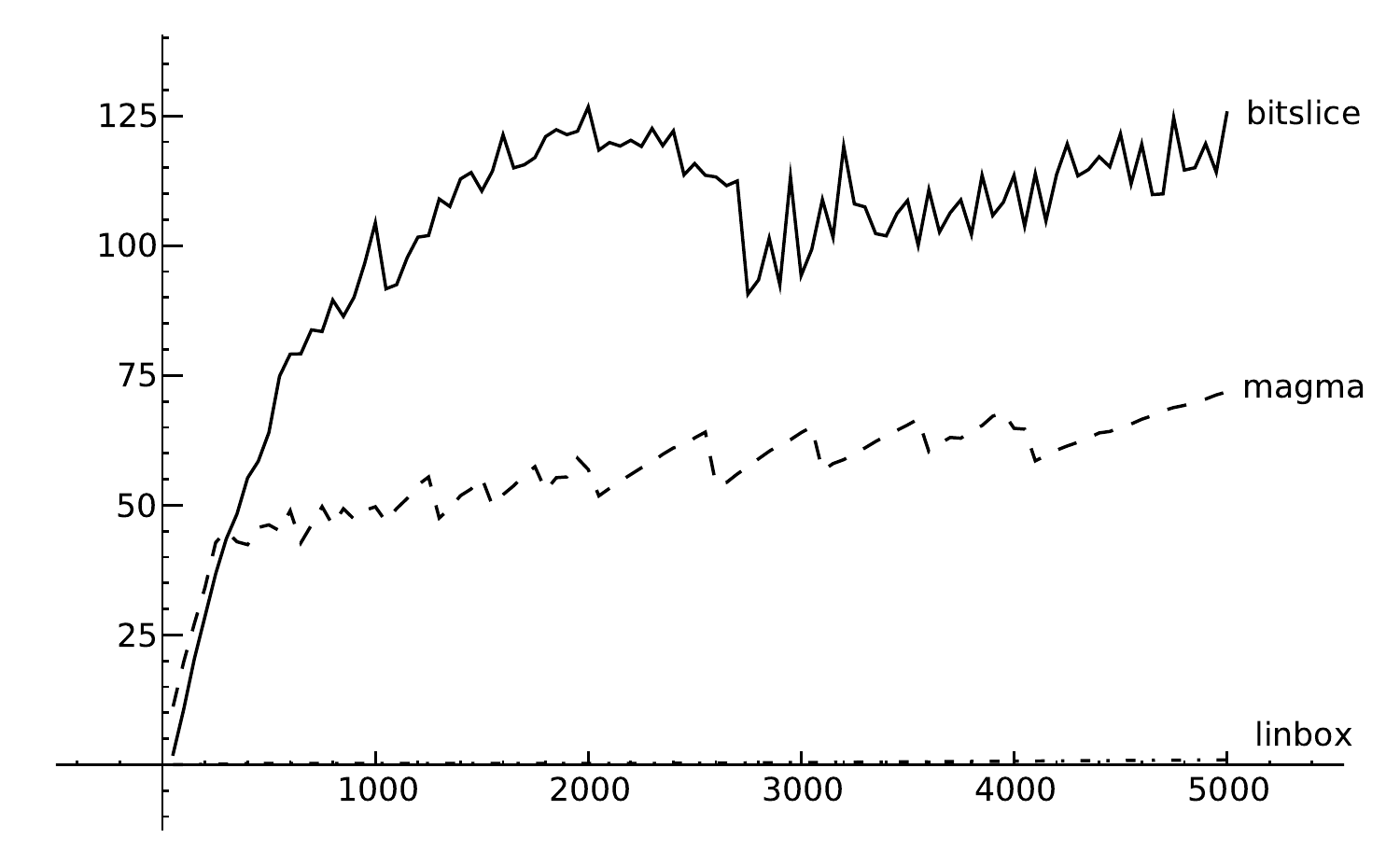}
 }
\caption{Effective GFFops ($2n^3/t$).}
\label{fig:ffops}
\end{figure}

We also compare our implementation with the FFLAS routine {\tt fgemm} for $\FF_p$ and using the Givaro Zech log representation for $\FF_{2^k}$, both part of LinBox \cite{linbox}. 
It should be noted that these are both much better suited to larger fields. 
Though we don't have an optimized implementation of the packing scheme proposed in \cite{DuFoSa08}, an effective upper bound can be placed by calculating the number of field elements that can be packed into a double and performing an the appropriately-sized floating point matrix multiply. 
A comparison for two specific fields can be seen in figure \ref{fig:ffops}. 
Though asymptotically faster algorithms are used, we normalize against the classical $O(n^3)$ to give an effective number of finite field operations per second (FFops). The jigsaw effect is due to the Method of Four Russians being sensitive to how close the matrix dimensions lie to a word boundary, and the sudden drop in efficiency for packed double is the transition from 4 elements per double to 3. 

Our implementation will be included in the open source math software Sage \cite{sage}.

\bibliographystyle{abbrv}
\bibliography{bitslice}

\begin{thebibliography}{10}

\bibitem{AUH}
A.~V. Aho, J.~E. Hopcroft, and J.~D. Ullman.
\newblock {\em The Design and Analysis of Computer Algorithms (Addison-Wesley
  Series in Computer Science and Information Processing)}.
\newblock {Addison Wesley}, January 1974.

\bibitem{m4ri}
M.~Albrecht and G.~Bard.
\newblock {\em The M4RI Library -- Version 20080901}.
\newblock The M4RI~Team, 2008.

\bibitem{M4RI-paper}
M.~Albrecht, G.~Bard, and W.~Hart.
\newblock Efficient multiplication of dense matrices over gf(2).
\newblock {\em CoRR}, abs/0811.1714, 2008.

\bibitem{M4RM}
V.~Arlazarov, E.~Dinic, M.~Kronrod, and I.~Faradzev.
\newblock On economical construction of the transitive closure of a directed
  graph.
\newblock {\em Dokl. Akad. Nauk.}, 194(11),194(11) (in Russian), English
  Translation in Soviet Math Dokl., 1970.

\bibitem{Bard06}
G.~V. Bard.
\newblock Accelerating cryptanalysis with the method of four russians.
\newblock Cryptology ePrint Archive, Report 2006/251, 2006.
\newblock http://eprint.iacr.org/.

\bibitem{Biham97}
E.~Biham.
\newblock A fast new {D}{E}{S} implementation in software.
\newblock In {\em Fast Software Encryption: 4th International Workshop, FSE'97,
  Haifa, Israel, January 1997. Proceedings}, volume 1267, pages 260--272.
  Springer-Verlag, 1997.

\bibitem{magma}
W.~Bosma, J.~Cannon, and C.~Playoust.
\newblock The {M}{A}{G}{M}{A} algebra system {I}: the user language.
\newblock {\em J. Symb. Comput.}, 24(3-4):235--265, 1997.

\bibitem{mod5modularity}
K.~Buzzard and W.~Stein.
\newblock A mod five approach to modularity of icosahedral galois
  representations.
\newblock {\em Pacific J. Math.}, 203 No. 2, 2002.

\bibitem{DuFoSa08}
J.-G. Dumas, L.~Fousse, and B.~Salvy.
\newblock Compressed modular matrix multiplication.
\newblock In M.~{Moreno Maza} and S.~M. Watt, editors, {\em Milestones in
  Computer Algebra (MICA 2008)}, May 2008.
\newblock Proceedings of a conference in honour of Keith Geddes' 60th birthday.
  Stonehaven Bay, Trinidad and Tobago, 1-3 May 2008.

\bibitem{REDQ}
J.-G. Dumas, L.~Fousse, and B.~Salvy.
\newblock Simultaneous modular reduction and kronecker substitution for small
  finite fields.
\newblock {\em CoRR}, abs/0809.0063, 2008.

\bibitem{FFLAS}
J.-G. Dumas, T.~Gautier, P.~Giorgi, and C.~Pernet.
\newblock Dense linear algebra over finite fields: the {F}{F}{L}{A}{S} and
  {F}{F}{P}{A}{C}{K} packages.
\newblock {\em CoRR}, abs/cs/0601133, 2006.

\bibitem{Karnaugh}
M.~Karnaugh.
\newblock The map method for synthesis of combinational logic circuits.
\newblock {\em AIEE Transactions Comm. Elec}, 72:593--599, 1953.

\bibitem{Kwan00}
M.~Kwan.
\newblock Reducing the gate count of bitslice {D}{E}{S}.
\newblock Cryptology ePrint Archive, Report 2000/051, 2000.
\newblock http://eprint.iacr.org/.

\bibitem{MaySauWan07}
J.~P. May, D.~Saunders, and Z.~Wan.
\newblock Efficient matrix rank computation with application to the study of
  strongly regular graphs.
\newblock In {\em ISSAC '07: Proceedings of the 2007 international symposium on
  Symbolic and algebraic computation}, pages 277--284, New York, NY, USA, 2007.
  ACM.

\bibitem{Pornin99}
T.~Pornin.
\newblock Automatic software optimization of block ciphers using bitslicing
  techniques.
\newblock In {\em Ecole Normale Superieure}, 1999.

\bibitem{Espresso}
R.~L. Rudell.
\newblock Multiple-valued logic minimization for pla synthesis.
\newblock Technical Report UCB/ERL M86/65, EECS Department, University of
  California, Berkeley, 1986.

\bibitem{LogicFriday}
{Sontrack}.
\newblock {\em {{L}ogic{F}riday} (Version 1.02)}, 2008.
\newblock {\tt http://sontrak.com}.

\bibitem{sage}
W.~Stein.
\newblock {\em {Sage}: {O}pen {S}ource {M}athematical {S}oftware}.
\newblock The Sage~Group, 2008.
\newblock {\tt http://www.sagemath.org}.

\bibitem{linbox}
{The LinBox Group}.
\newblock {\em {{L}in{B}ox}: Exact Linear Algebra with Dense and BlackBox
  Matrices, ({V}ersion 1.1.5)}, 2008.
\newblock {\tt http://www.linalg.org}.

\bibitem{WengQuiWangXiang07}
G.~Weng, W.~Qiu, Z.~Wang, and Q.~Xiang.
\newblock Pseudo-paley graphs and skew hadamard difference sets from
  presemifields.
\newblock {\em Des. Codes Cryptography}, 44(1-3):49--62, 2007.

\end{thebibliography}

\end{document}